\documentclass[12pt]{iopart}

\usepackage{graphicx}
\usepackage{cite}

\begin{document}

\title{Plateau-insulator transition in graphene}
\author{M Amado$^{1,2}$, E Diez$^1$, D L\'opez-Romero$^3$, F Rossella$^4$, J M Caridad$^{1,4}$, F Dionigi$^4$, V Bellani$^4$, and D K Maude$^5$.}

\address{$^1$ Laboratorio de Bajas Temperaturas, Universidad de Salamanca, E-37008 Salamanca, Spain}

\address{$^2$ GISC-QNS,Dpto. de F\'{\i}sica de Materiales, Universidad Complutense, E-28040 Madrid, Spain}

\address{$^3$ CT-ISOM, Universidad Polit\'ecnica de Madrid, E-28040 Madrid, Spain}

\address{$^4$ Dipartimento di Fisica ``A. Volta'' and CNISM, Universit\`{a} degli studi di Pavia, I-27100 Pavia, Italy}

\address{$^5$ Laboratoire National des Champs Magn\'etiques Intenses, F-38042 Grenoble, France}

\ead{marioam@fis.ucm.es}

\begin{abstract}
We investigate the quantum Hall effect (QHE) in a graphene sample with Hall-bar geometry close to the Dirac point at high magnetic fields up to 28~T. We have discovered a plateau-insulator (PI) quantum phase transition passing from the last plateau for the integer QHE in graphene to an insulator regime $\nu=-2 \rightarrow \nu=0$. The analysis of the temperature dependence of the longitudinal resistance gives a value for the critical exponent associated to the transition equal to $\kappa=0.58\pm 0.03$.

\end{abstract}


\submitto{\NJP}

\maketitle

\date{\today}

\section{Introduction}
The recent discovery of graphene~\cite{novoselov2004}, a flat monolayer of carbon atoms tightly packed into a two-dimensional (2D) honeycomb lattice, has attracted great attention, in particular after the experimental observation of a non standard sequence of integer quantum Hall features~\cite{novoselov2005,stormer2005}. Graphene is the first truly 2D system available for investigation by solid state physicists. The unusual sequence of integer quantum Hall features is related to the quasi-relativistic nature of the charge carriers due to the particular graphene band structure near the Dirac point~\cite{guinea2}. In graphene the conventional integer quantum Hall quantization for the conductivity $\sigma_{xy}$ is shifted by a half integer~\cite{guinea2, gusynin1} due to the half-filling of the $n=0$ Landau level compared to the other levels:
\begin{equation}
\label{sigmahall} \sigma_{xy}= \frac{4 e^2}{h} \left(\pm n+ \frac{1}{2}\right)=\frac{\nu e^2}{h} \,.
\end{equation}
At very high magnetic fields and with high quality samples more quantum Hall plateaus not in the sequence of
Eq.(\ref{sigmahall}) have been observed~\cite{Zhang,Abanin,Jiang,Ong1, Giesbers}; these include filling factors
$\nu=0,\pm1,\pm4$ and the fractional quantum Hall effect has been also observed in suspended graphene samples~\cite{Bolotin,Du} with an insulating phase in the state $\nu=0$~\cite{Du}. These new plateaus cannot be understood only by using the Landau quantization and their origin is currently a topic under a considerable debate. Many theories attribute the origin of these new plateaus to some spontaneously broken symmetry of the system driven by electron-electron interactions (see Ref.~\cite{yang1} and references therein). Edge states could also play a key role and their relevance for the dynamics in graphene~\cite{Abanin,Castro,Abanin2,Brey,fertig,shimshoni1} has been invoked to explain a variety of magneto-transport experiments near $\nu=0$. Different measurements have demonstrated that the longitudinal resistance may either decrease~\cite{Abanin} or increase~\cite{Ong1} at $\nu \sim 0$ with decreasing temperature in samples that appear to be quite similar, fueling the debate concerning the existence and origin of an insulator phase near the Dirac point at high magnetic fields. This insulator phase has been attributed to field-induced spin-density waves in non-suspended graphene samples~\cite{Jung}. Our data shows clearly a transition to an insulator phase at high magnetic fields being consistent
with the results by Checkelsky \emph{et al}.~\cite{Ong1} and Giesbers \emph{et al}.~\cite{Giesbers} but in a rather different regime, away from the Dirac point.

In the framework of the scaling theory of the integer quantum Hall effect (IQHE),
plateau-plateau (PP) and plateau-insulator (PI) transitions are interpreted as quantum phase transitions with an associated
universal critical exponent $\kappa$. The underlying physics is the Anderson localization-delocalization quantum phase
transition~\cite{tsui}. In this work we report experimental evidence for a PI transition between states at
filling factors $\nu=-2$ and $\nu=0$. We have obtained a value for the scaling exponent equal to $\kappa=0.58\pm0.03$, to our knowledge, the first evidence of this exponent in a truly 2D system. The value we found for $\kappa$ is very similar to the values measured for this exponent for the plateau-insulator transition in  standard quasi two-dimensional electron gases (2DEGs) in InGaAs/InP and InGaAs/GaAs quantum wells (see~\cite{Schaijk,deLang,ponomarenko1,Hilke1,Shahar1,Pruisken} and~\cite{ponomarenkoTH} for a complete review).

\section{Sample and magnetotransport measurements}
\begin{figure}[ht]
 %\vspace{5mm}
 \begin{minipage}{.5\linewidth}
 \includegraphics[width=70mm,clip]{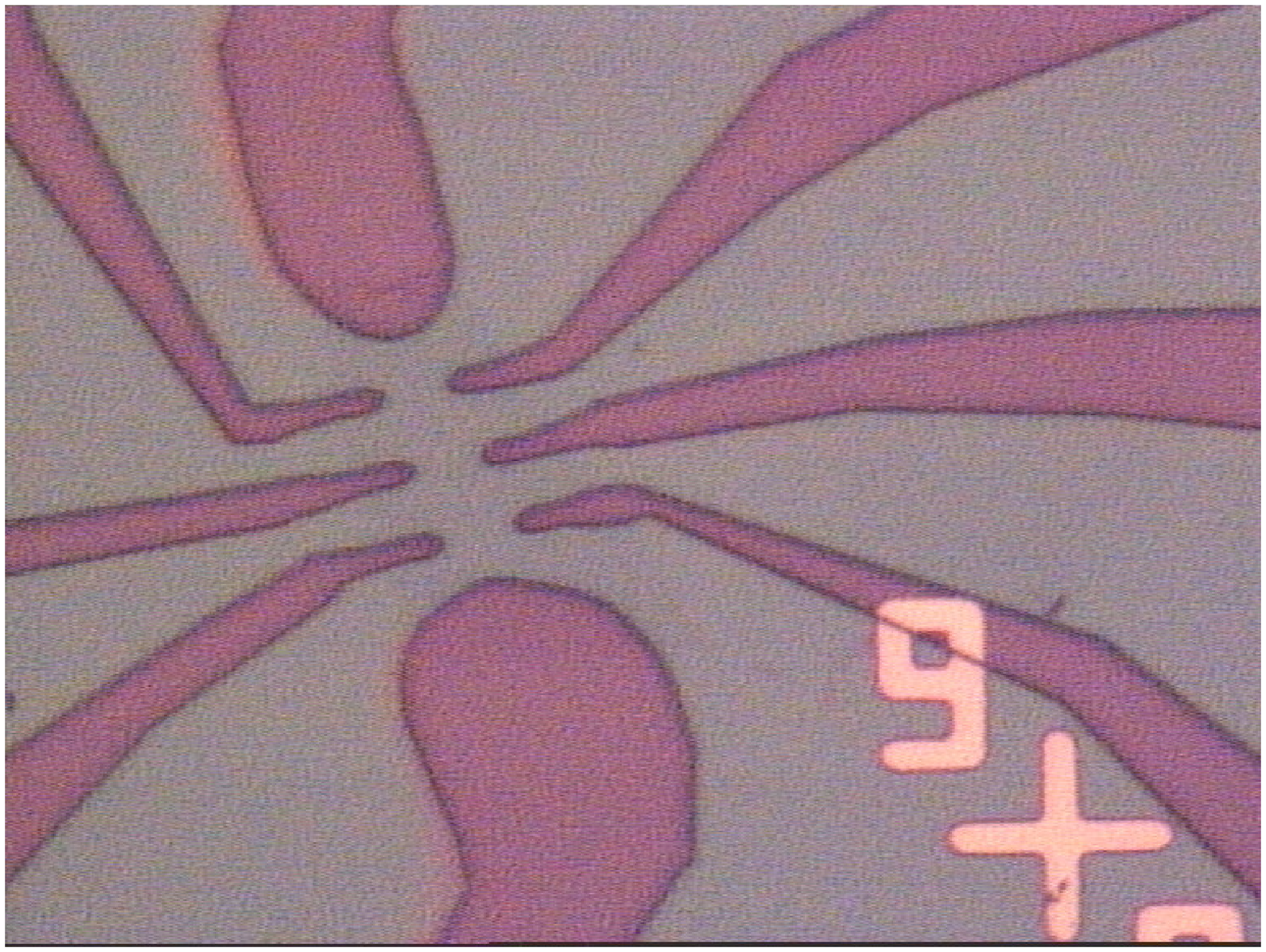}
 \end{minipage}%
\hspace{2mm}
\begin{minipage}{.5\linewidth}
 \includegraphics[width=70mm,clip]{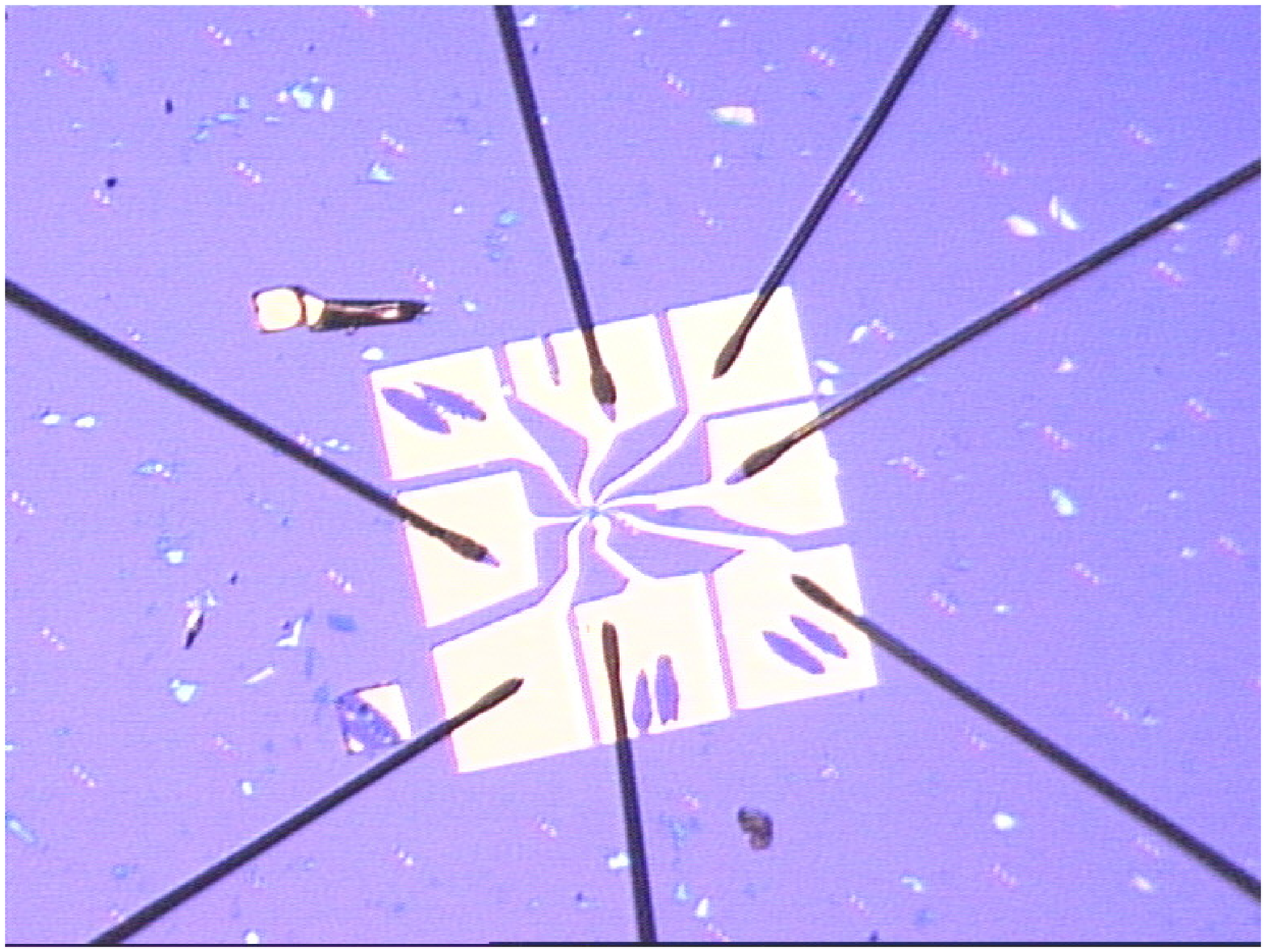}
 \end{minipage}%
\caption{\textit{The left image shows the sample with the Hall bar covered by PMMA resist before the deposition of the contacts while the right one the contacted sample with the Ti/Au already deposited.}}
\label{sample}
\end{figure}

We have used monolayer graphene flakes obtained by peeling graphite onto a Si wafer with a $300$~nm SiO$_2$ top layer. They were processed in the CT-ISOM (Central de Tecnolog\'{i}a-Instituto de Sistemas Optoelectr\'{o}nicos y Microtecnolog\'{i}a) clean-room facilities,
depositing $50/500 \AA$ Ti/Au contacts in a Hall-bar geometry by using e-beam nanolithography as can be seen in Fig.~\ref{sample}).

 We made 12 hall-bars in different monolayer graphene flakes and we characterize them by magnetotransport measurements in our laboratory in Salamanca. We selected the sample S2601f4 that exhibited a high degree of homogeneity and the highest mobility. The Hall mobility ($\mu$) for this sample ranges from $\mu=4.5 \times 10^3 cm^2 V^{-1} s^{-1}$ at a gate voltage $V_{gate}=17$V  far away from the Dirac point to $\mu=1.3 \times 10^4 cm^2 V^{-1} s^{-1}$ at $V_{gate}=5$V and being  $4.2 \times 10^4 cm^2 V^{-1} s^{-1}$ nearby the Dirac point at $V_{gate}=1$V.

\begin{figure}
\centerline{\includegraphics[width = 9cm]{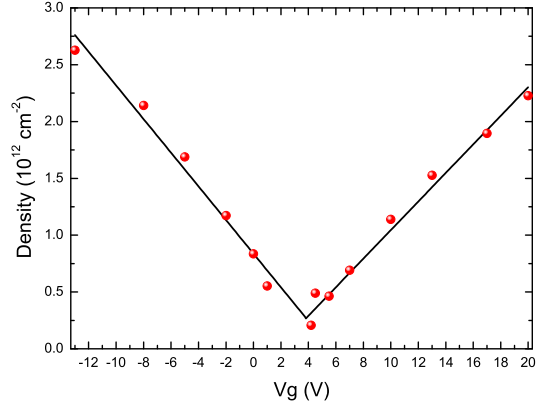}}
\caption{Density in the electron an hole regime measured at $4.2$K (red dots) and interpolated function of the density (black line) for the sample prior to the annealing. As it can be seen, the Dirac point was at $3.8$V while it was displaced to $-3$V during the transport experiments.} \label{density-vs-Vg}
\end{figure}

We have performed magneto-transport experiments in this sample up to $28$~T at the Grenoble High Magnetic Field Laboratory using a $20$~MW resistive magnet. On a first approach we measured the density in a dilution fridge  and the position of the Dirac point that was found at $3.8$V at $T=200$mK .
The carrier concentration (electron and hole density in $10^{12} cm^{-2}$) as a function of the gate voltage $V_{gate}$ at $B=0$ and $4.2$K is shown in Fig.~\ref{density-vs-Vg}. The gate voltage can be easily transformed into the carrier concentration given by $n=\frac{C}{A}\frac{(V_{gate}-V_{Dirac})}{e}$ with $C/A=2.3\times10^{-4}Fm^{-2}$ in the hole regime and $C/A=2.0\times10^{-4}Fm^{-2}$ in the electron one (note the slight asymmetry in the obtained fitting curve) where $C$ is the capacitance of the device ($C\sim0.02pF$) and $A$ its area ($A=107\mu m^2$).

\begin{figure}[ht]
 %\vspace{5mm}
 \begin{minipage}{.5\linewidth}
 \includegraphics[width=80mm,clip]{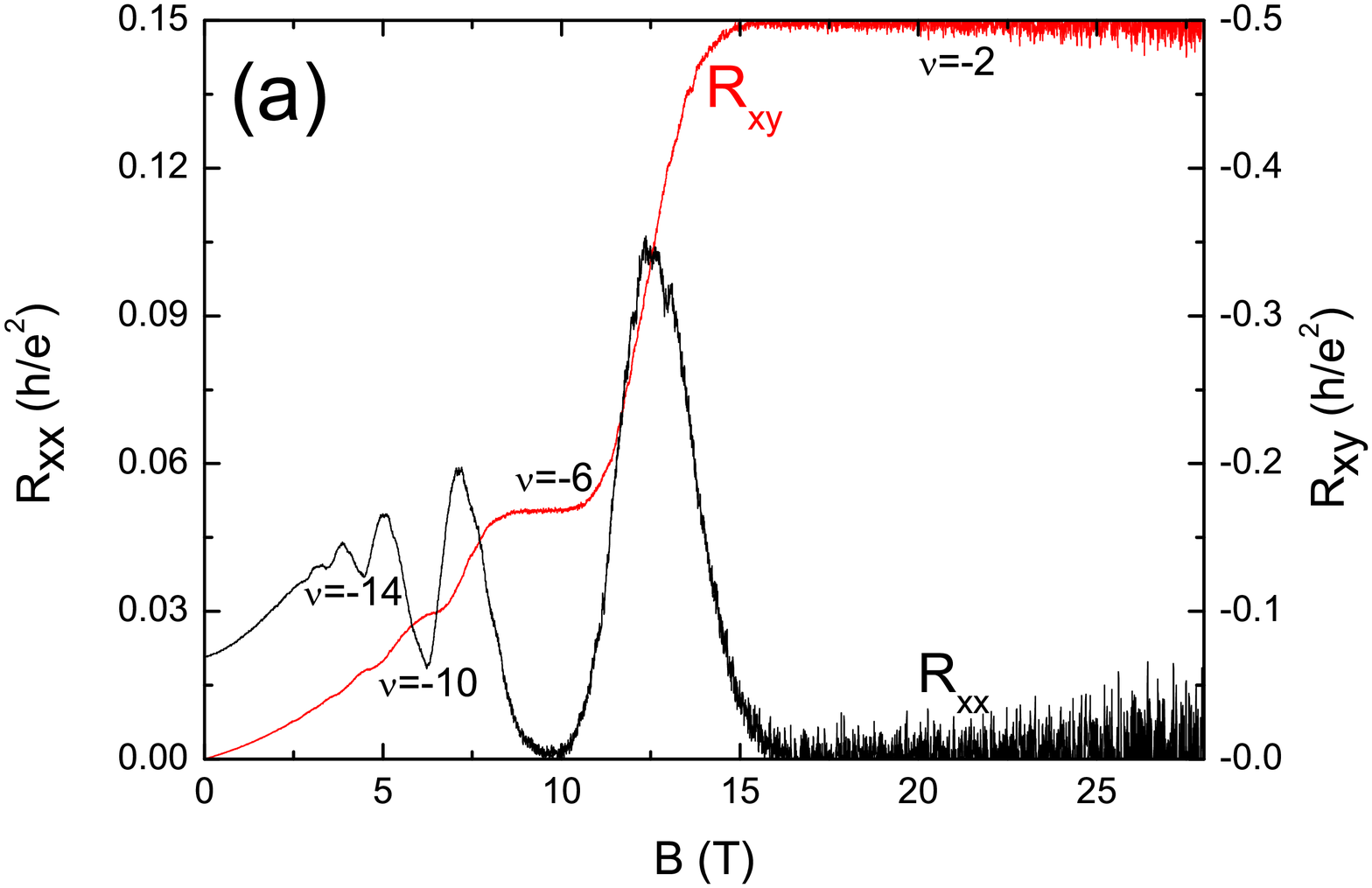}
 \end{minipage}%
\hspace{2mm}
\begin{minipage}{.5\linewidth}
 \includegraphics[width=80mm,clip]{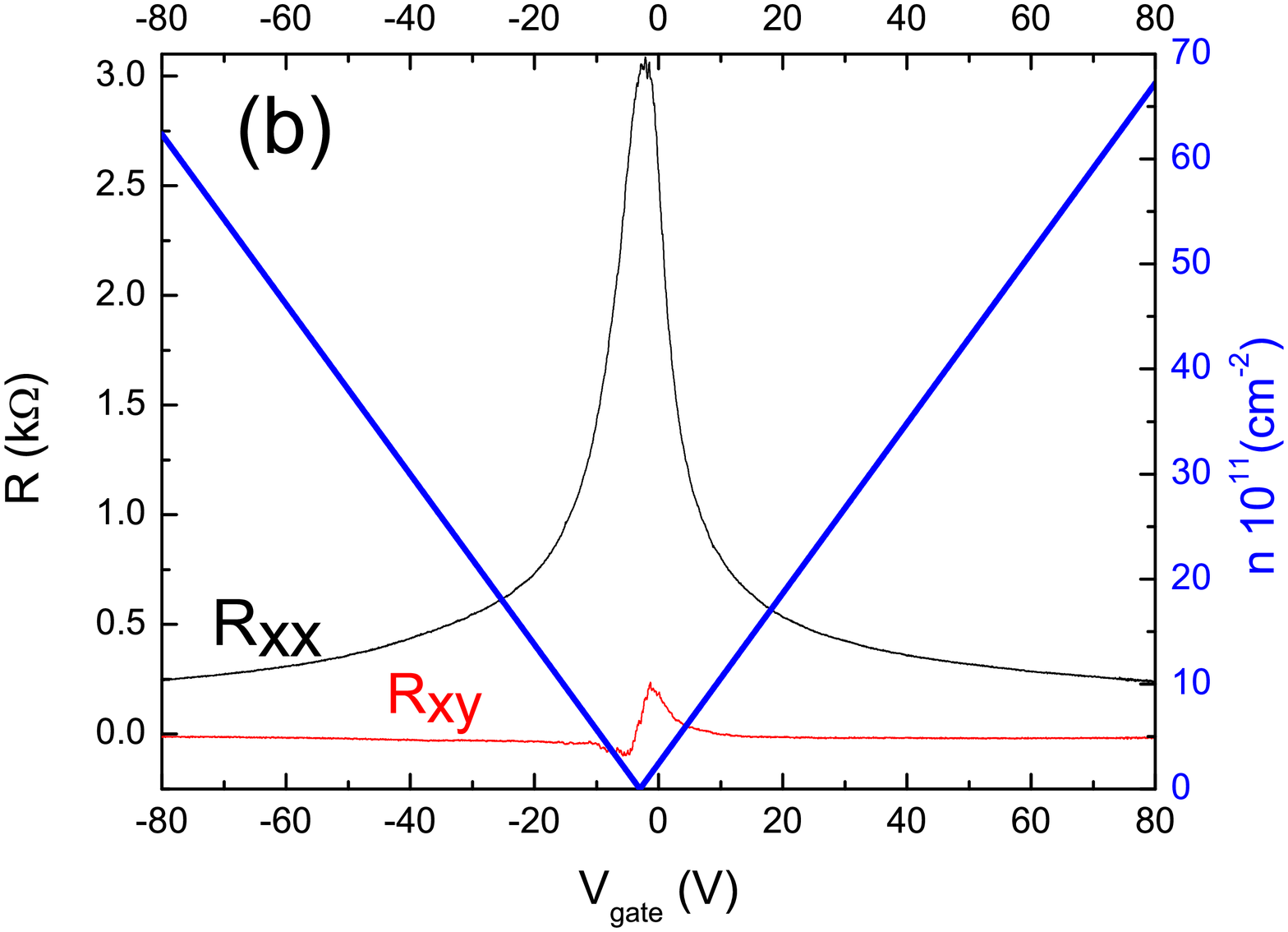}
 \end{minipage}
\caption{\textit{(a) $R_{xx}$ (black line) and $R_{xy}$ (red line) as a function of the magnetic field in the hole-like region measured at $4.2$ K for $V_{gate}=-20\,$V. In this case the density is $n_{2D}=1.4 \times 10^{12}\,cm^{-2}$ and the mobility  $\mu=4.5 \times 10^3 cm^2 V^{-1} s^{-1}$.
(b) $R_{xx}$ and $R_{xy}$ at $B=0$T as a function of gate voltage, the Dirac point is centered at $V_{Dirac}=-3$ V. The interpolated density of carriers measured at $4.2$K via the Hall slope (in $10^{11} cm^{-2}$) is also showed for clarity }}
\label{R_vs_B_Vg}
\end{figure}

The sample was then annealed $90$ minutes in a low pressure He exchange gas atmosphere at 100 $^o$C  in-situ prior to insertion into the $^4$He cryostat. We have observed a displacement of the Dirac point to $-3$V (see Fig.\ref{R_vs_B_Vg}(b) becoming it slightly narrower. Four probe measurements were carried out using standard ac lock-in techniques with excitation currents of $1-10$~nA and frequencies of $1.3-13$~Hz.

\section{Results and discussion}

In figure {\ref{R_vs_B_Vg} (a)} the variation of the Hall ($R_{xy}$) and longitudinal resistance ($R_{xx}$) as a function of the magnetic field are showed. The temperature was $4.2$~K and the gate voltage $V_{gate}=-20\,$V. In such a configuration we had the Fermi energy far away from the Dirac point with several Landau levels below it. At this gate voltage and keeping the temperature constant at $4.2$K we extracted the hole density from the Hall measurements, being $n_{2D}=1.4 \times 10^{12} cm^{-2}$  with $\mu = 4.5 \times 10^3 cm^2 V^{-1} s^{-1}$. The Hall resistance displays plateaus with the expected sequence for single-layer graphene in accordance with Eq. (\ref{sigmahall}). Figure {\ref{R_vs_B_Vg} (b)} shows $R_{xx}$, and $R_{xy}$ as a function of the gate voltage without applying an external magnetic field . $R_{xy}$ at B=0 shows a peak close to the Dirac point because $R_{xx}$ is much larger than $R_{xy}$ and even with a small Hall-contact misalignment $R_{xx}$ contribution is dominant. The Dirac point is centered at $V_{Dirac}=-3\,$ V remaining invariable in the following measurements showed in this paper. The interpolated density of carriers (obtained at $4.2$K using the same technique as in Fig.~\ref{density-vs-Vg}) is shown to extract easily its value compared with the position of the Dirac point.

\begin{figure}[ht]
 %\vspace{5mm}
 \begin{minipage}{.5\linewidth}
 \includegraphics[width=80mm,clip]{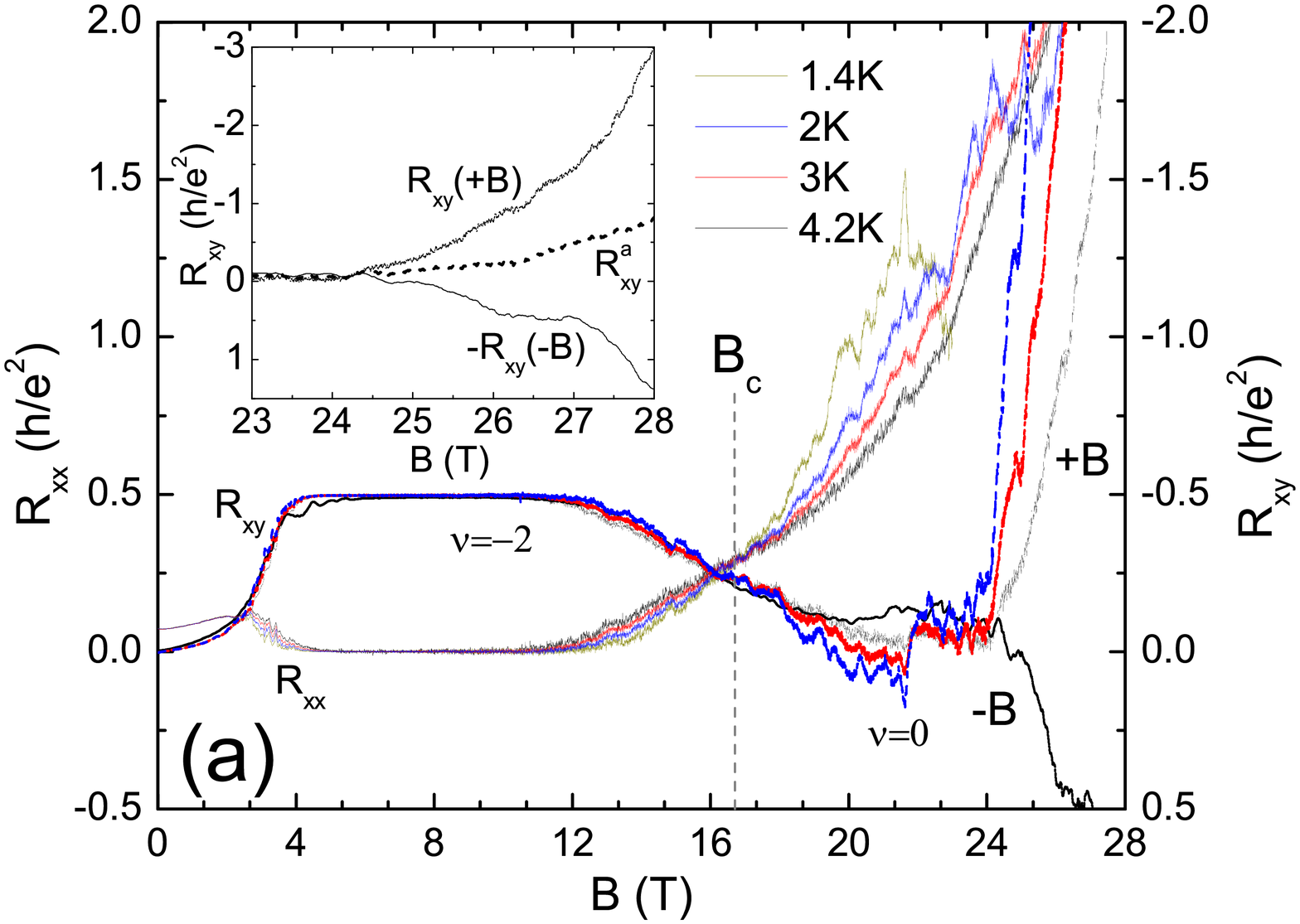}
 \end{minipage}%
\hspace{2mm}
\begin{minipage}{.5\linewidth}
 \includegraphics[width=80mm,clip]{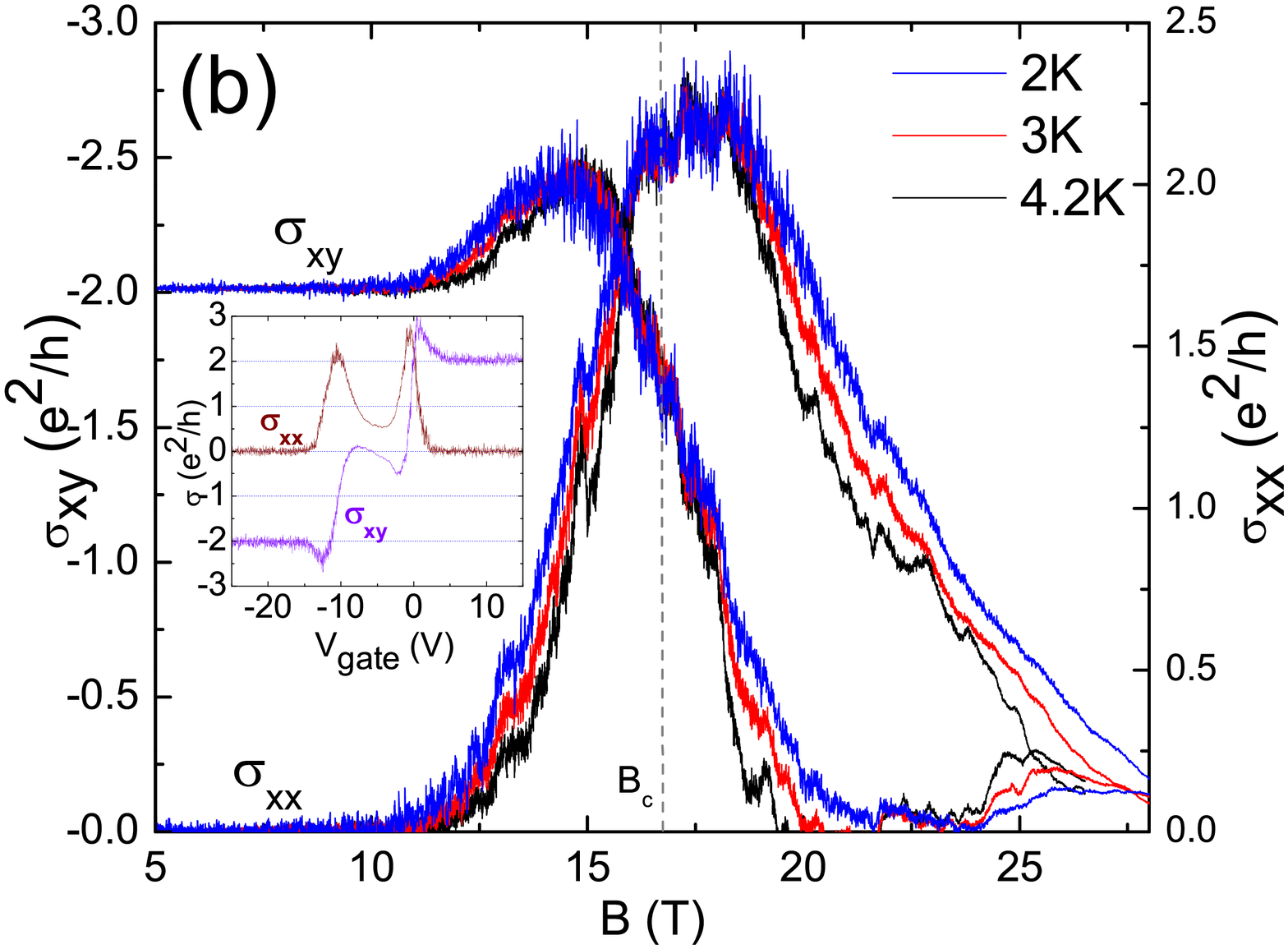}
 \end{minipage}%
\caption{(a) Longitudinal ($R_{xx}$) and Hall ($R_{xy}$) resistances at $T = 1.4$, $2, 3$ and $4.2$~K and (b) the conductivities $\sigma_{xx}$ and
$\sigma_{xy}$ (b) as a function of B for $T = 2, 3$ and $4.2$~K with a $V_{gate}= -8\,$V. The hole density is $n=4.1 \times 10^{11}\,cm^{-2}$ and the  mobility is $\mu=1.3 \times 10^4 cm^2 V^{-1} s^{-1}$ at $T=4.2$K. In (a) the inset shows the Hall resistance for $+B$ and $-B$ in solid lines and $R_{xy}^a =\frac{1}{2}[R_{xy}(+B) - R_{xy}(-B)]$ in dashed line. In the inset of (b) we show the longitudinal $\sigma_{xx}$ and Hall $\sigma_{xy}$ conductivities as a function of the gate voltage at $B=24$~T and $4.2$~K. $\sigma_{xx}$ show a double peak structure, indicating the splitting of the fourfold degeneracy. Meanwhile, the Hall conductivity shows a central plateau on which $\sigma_{xy}= 0$, with two lateral plateaus corresponding to $\sigma_{xy} = \pm 2e^2/h$.} \label{conductivities}
\end{figure}

Then we moved the gate voltage to $V_{gate}= -8\,$V, close enough to the Dirac point to have only the lowest Landau level ($\nu=-2$) below the Fermi energy. Applying  a magnetic field above $15$T we observed a transition to a new quantum Hall plateau ($\nu=0$) not in the sequence of Eq.(\ref{sigmahall}) as previously observed by  other groups~\cite{Zhang,Abanin,Jiang,Ong1}. We focused our study to the temperature dependence of this transition. In Fig. \ref{conductivities}(a) we show $R_{xx}$ and $R_{xy}$ and in Figure \ref{conductivities}(b) the conductivities $\sigma_{xx}$ and $\sigma_{xy}$ (longitudinal and Hall) as a function of the magnetic field from $1.4$ to $4.2$ K, with $V_{gate}= -8\,$V. The hole density was $n=4.1 \times 10^{11}\,cm^{-2}$ and the  mobility $\mu=1.3 \times 10^4 cm^2 V^{-1} s^{-1}$. We observed a transition from the $\nu=-2$ to $\nu=0$ state. The longitudinal resistance grown exponentially and appeared to exhibit a temperature independent critical field at $B_c=16.7\,$T.
The Hall resistance has also a a temperature independent critical point at $B_c=16.7\,$T and tended to the expected quantized value $\nu=0$ reaching it at $B>18$T for at least $5$T. Unfortunately, the diverging behavior of $R_{xx}$ seriously hindered the determination of the Hall resistance $R_{xy}$, as even a small Hall-contact misalignment would result as the dominant contribution from $R_{xx}$. The traditional way to overcome this problem is by using the $B$ symmetry of the resistance components: R$_{xx}$ is expected to be symmetric when applying external B, while R$_{xy}$ to be anti-symmetric~\cite{Hilke1}. The corrected $R_{xy}$ is obtained, by an anti-symmetrization of the measured $R_{xy}(+B)$ and $-R_{xy}(-B)$: $R_{xy}^a=[\frac{1}{2}(R_{xy}(+B) - R_{xy}(-B))]$  measuring with reversed external magnetic field configurations $(+B)$ and $(-B)$ (both perpendicular to the 2DEG, being (+B) the standard configuration we have used in the work and (-B) the reversed magnetic field configuration, see the inset in Figure \ref{conductivities} for clarity). In practice, because the admixture is a result of both a misalignment of contacts and a nonuniform current distribution in the sample, $R_{xx}$ itself is not entirely B-symmetric, limiting the accuracy to which $R_{xy}$ can be determined. Nevertheless, the corrected Hall resistance remained quantized deep into the insulating phase.

Our observations are likely to be understandable in terms of the usual picture for QH PI transitions.
In a somewhat simplified picture of the standard QHE in 2DEGs, the transitions between plateaus take place at $\nu=\frac{3}{2}, \frac{5}{2}, \frac{7}{2},$ etc. A special case is the transition from the last Hall plateau at $\nu=1$ that occurs at $\nu=\frac{1}{2}$ where the 2DEG, instead of undergoing a transition between two plateaus, becomes an insulator. In the case of the PI transition in graphene from $\nu=-2$ to $\nu=0$ the expected filling factor at the transition is $\nu=-1$, almost the same that we have obtained in our experiment $\nu(B_c)=-1.015$ confirming our interpretation of this transition.

\begin{figure}
\centerline{\includegraphics[width = 9cm]{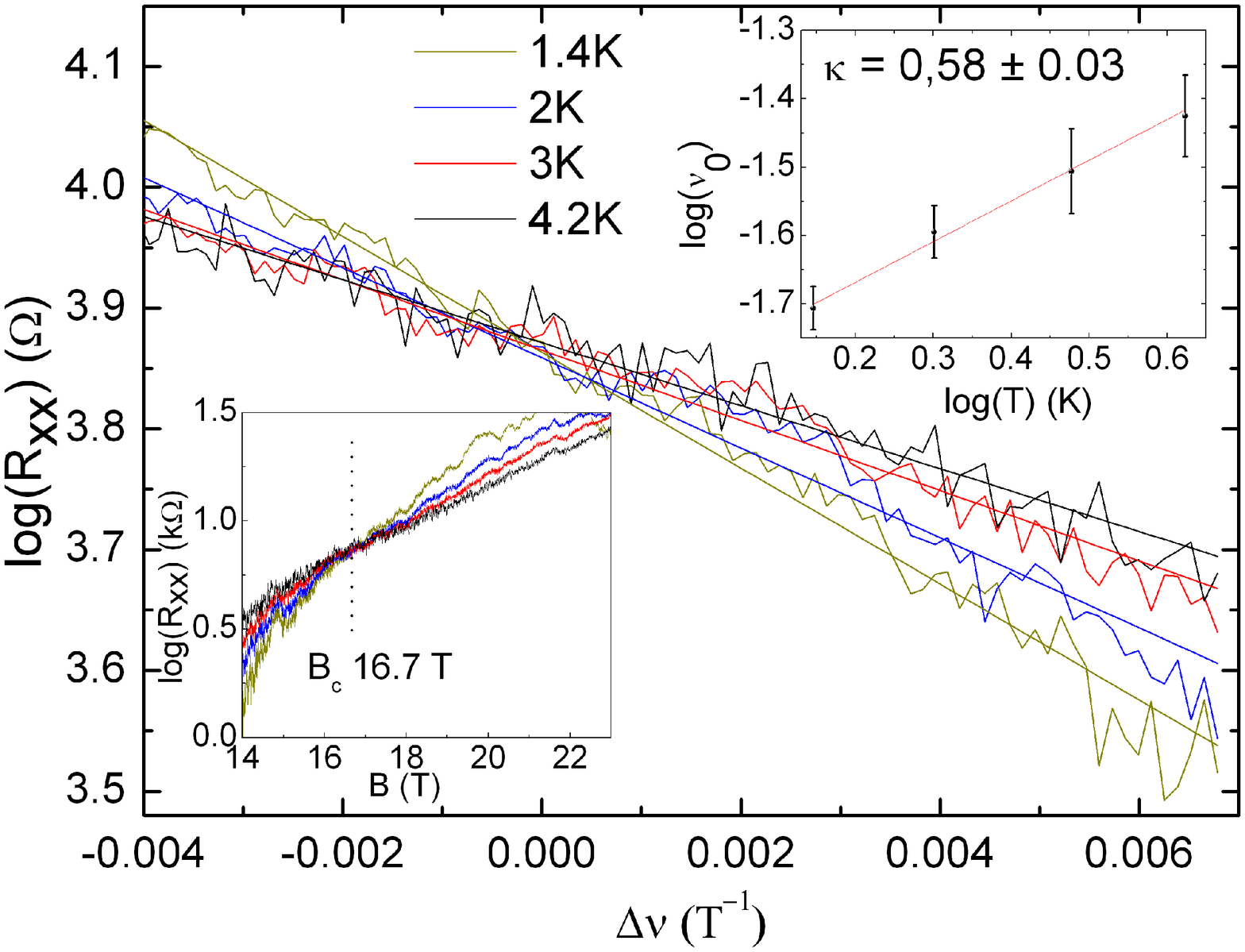}}
\caption{Longitudinal resistance ($R_{xx}$) as a function of $\Delta \nu$ with $\Delta \nu = 1/B$ - $1/B_c$ and
$B_c=16.7\,$ T at four different temperatures. In the lower inset we show $log(R_{xx})$ as a function of B. Using the standard scaling procedure we
obtain $\nu_0$ by fitting our data to Eq.(\ref{scaling1}) and the results are plotted in the upper inset. We obtain a critical exponent $\kappa=0.58\pm0.03$.
} \label{fitting}
\end{figure}

In this study of the PI transition we kept constant the carrier density (constant gate potential) then tuning $1/B$ was equivalent to tune the filling factor. The experimental characterization of quantum Hall plateau-plateau and plateau-insulator quantum phase transitions relies on the behavior of the system as a function of T for magnetic fields sufficiently close to a transition point. In the vicinity of this transition point the longitudinal resistance is expected to follow the
empirical law~\cite{Shahar1,Pruisken}:
\begin{equation}
\label{scaling1} R_{xx}= \exp[-\Delta \nu/\nu_0(T)]\,,
\end{equation}
with $\Delta \nu = 1/B - 1/B_c$ ($\nu$ here should not be confused with the Landau level filling factor) and $B_c$ is
the temperature independent critical field set as $16.7$ T as said before [left inset in Fig. (\ref{fitting})]. Fitting to Eq. (\ref{scaling1}) the experimental $R_{xx}$ measured at four
different temperatures we can obtain the  associated  critical exponent $\kappa$:$ \nu_{0}\propto T^{\kappa}$.

This approach has been used to study the plateau-insulator transition in a great variety of quasi 2D systems. The seminal magneto-transport experiments of Tsui \emph{et al}.~\cite{tsui} in InGaAs/InP samples found a value for $\kappa$ of $0.42$ in PP transitions. For a PI transition, although there remained some controversy, after two decades of experiments the most commonly measured value is $\kappa=0.57$. This value has been found for a vast range of samples and materials, in agreement also with numerical simulations~\cite{Schaijk,deLang,ponomarenko1,Hilke1,Shahar1,Pruisken,ponomarenkoTH}

In Fig.~\ref{fitting} we plot $R_{xx}$ as a function of $\Delta \nu$. We have not drawn possible maximum-minimum lines that can be extracted from upper and lower-bound values because it will reflect the strenght of the white noise, observed in magnetotransport experiments at very high magnetic fields (see~\cite{Giesbers} as an example).We have fitted our data to the empirical law (\ref{scaling1}) with a typical least squares fit, and the goodness of single fit was testified by the fact that $R^2$ was about $0.97$. We noted that the linear fit reproduces the behavior of the longitudinal resistance, and this indicates that the noise do not affect too much to the least squares accuracy. From these fits we obtained $\nu_0$ that
are plotted in the right inset of Fig. \ref{fitting} as a function of T. The exponent $\kappa$ was then obtained from $ \nu_{0} \propto T^{\kappa}$ being the best fit  $\kappa=0.58\pm0.03$ (with $R^2=0.99$). To obtain the overall incertitude we introduced the error of each single fit and temperature stability ($10$mK).

We firmly believe that this is a clear evidence of the existence of a quantum Hall insulator in graphene in the $\nu=0$ state away enough from the Dirac point and that the transition from the last quantum Hall liquid
($\nu=\pm2$) is the same (i.e belongs to the same universality class) as the well-known PI transition in conventional quasi 2D systems.

 Previous experiments in the zero-energy state in graphene in the presence of high magnetic fields were focused on the dependence with the gate voltage~\cite{Abanin,Jiang,Ong1} (then for a fixed magnetic field) and close to the Dirac point. We observed that using magnetic field as the driving parameter rather than the gate voltage we can obtain a more reliable temperature independent critical point. In previous studies the $\nu=0$ quantum Hall state does not show a clear plateau in $R_{xy}$, being only visible as a plateau in the Hall conductance~\cite{Jiang,Ong1}. Our measurements, although fluctuating, show a plateau in the Hall resistance at $\nu=0$. As we discuss above, the diverging behavior of the Hall resistance for magnetic fields above $24~T$ (see Fig.~\ref{conductivities}) is most likely due to Hall-contact misalignment.
Unfortunately our cryostat allowed us a small range of temperatures that should be
enlarged in future experiments. On the other hand, as it has been seen in previous experiments close to the Dirac point, the measurements shown large fluctuations and the nature of this behavior is not clearly understood~\cite{Ong1}. Of course we checked our results and in particularly our fittings are stable and reproducible and that our main results are not affected by these fluctuations.

Checkelsky \emph{et al.} reported temperature dependent transport measurements~\cite{Ong1} where $R_{xx}$ exhibits a magnetic field-induced crossover to a state with a very large resistance, as in our measurements. However, they observed that for a fixed magnetic field the resistance was almost temperature independent below $2~K$. This behavior, which appears to exclude a PI transition, was interpreted in terms of the symmetry breaking effect of the disorder potential which may favor long-range order at {\em finite} temperature with a Kosterlitz-Thouless (KT) transition~\cite{KT,Ong1}. In their case, the applied voltage was set extremely close to the Dirac point or even on it while in our case was $5$V away from it (note the big difference in terms of carrier density and mobility). They also observed how the divergence of $R_{xx}$ was shifted to higher fields when the Dirac point moved away from $V_g=0$V noting that it was important to choose samples with $|V_g|<1$V to investigate the intrinsic properties of the Dirac point, that is not our scenario.

We want to stress that our results are not in contradiction with the KT-like transition but rather probes a different regime. Away from the Dirac point, hence, when the magnetic field is varied in its strong regime (above 10T the Zeeman splitting seems to be strong enough) the Fermi level is tuned through the first (spin-resolved) Landau level. Then the features reported in this work can be easily understood in term of a transition dominated by percolation of the extended state.

The transition to an insulator reported in Ref.~\cite{Ong1} occurs when the doping is set exactly at the Dirac point. Then, regardless of how strong B is (as long as the spin is resolved), the Fermi level in the bulk is always precisely in the middle of the Zeeman gap. Due to particle-hole symmetry, the filling factor as well $\sigma_{xy}$ and $R_{xy}$ vanish to $0$. In this situation, the only gapless states that can in principle give rise to non-trivial dissipative transport are the "helical" edge states discussed recently by Shimshoni \emph{et al}.~\cite{shimshoni1}. In addition, we do not observe any trace of saturation over the range of temperatures investigated as in previous references~\cite{Ong1}. Nevertheless, we would like to stress that a temperature independent resistance has been observed in several experimental studies of PI quantum phase transitions. Shahar et al.~\cite{Shahar1} reported for the first time this type of behavior in a number of different 2D systems and although is not well understood it is usually related to some kind of long-range interactions~\cite{Pruisken} or to finite size effects~\cite{Wanli}. In fact, finite size effects can be of particular relevance here due to the small size of graphene samples. Also, the observed fluctuations in the measurements of $R_{xx}$ and $R_{xy}$ pointed out in our data could be at least partially related to mesoscopic effects~\cite{Peled1}.

We therefore conclude that our data shows clearly a PI transition from $\nu=-2$ to $\nu=0$ most likely due to the presence of Anderson localized states. We stress, however, that this do not exclude that such a Kosterlitz-Thouless transition could exist for higher magnetic fields or when the doping was precisely set at the Dirac point being probable that both scenarios could coexist.

\section{Summary and conclusions}

Summarizing, we have studied the quantum Hall effect in graphene and in particular the $\nu=-2$ to $\nu=0$ transition. We observed a PI transition when the magnetic field is tuned across a critical field $B_c=16.7\,$T. Both the Hall conductivity and the Hall resistance remain quantized deep into the insulating phase suggesting that we are in the quantized Hall insulator regime, the same that was observed for the PI transition in low mobility 2D systems. Using the standard scaling theory analysis we have obtained the critical exponent for this transition $\kappa=0.58\pm0.03$, in agreement with the common value observed for this exponent in conventional 2D systems. This evidences support the identification of $\nu=0$ in graphene as an insulating phase rather than a Quantum Hall state. To our knowledge this is the first measurement of this exponent in graphene and therefore the first measurement in a truly 2D system. Therefore two distinct PI transitions exist near the Dirac point in graphene, namely a conventional QH-insulator transition as the observed in our experiment and the recently observed Kosterlitz-Thouless transition by Checkelsky \emph{et al}.~\cite{Ong1}. Further studies at lower temperatures and higher magnetic fields are needed to confirm our results in a larger range of temperatures and magnetic fields, and in particular to try to find the conditions where both transitions could be observed at the same time.

\section*{Acknowledgments}
We are grateful to J. M. Cerver\'o, F. Dom\'{\i}nguez-Adame, E. Shimshoni and N. Azarova for the critical reading of the manuscript and for the enlighten suggestions, and to L. Brey and F. Guinea for useful discussions. This work was supported by the following projects: Cariplo Foundation QUANTDEV, MEC FIS2006-00716, PPT-31000-2008-3, MOSAICO, JCyL SA052A07, and European Union CTA-228043-EuroMagNET II Programme. We thank to P. Blake and M. M. Sanz  for  kind help during sample processing.
% \end{acknowledgments}

%\end{multicols}

\end{document}